\documentclass{elsart}

 \usepackage{graphicx}

\usepackage{amssymb}

\newcommand{\fm}{$\mathrm{fm}$}
\newcommand{\gev}{$\mathrm{(GeV/c)}^{2}$}
\newcommand{\qsq}{$\mathrm{Q^2}$}

\newcommand{\een}{$(e,e'n)$}   

\newcommand{\deep}{$D(e,e'p)$}
\newcommand{\deen}{$D(e,e'n)$}

\newcommand{\np}{$H(n,p)n$}
\newcommand{\gen}{$G_{en}$}
\newcommand{\gmn}{$G_{mn}$}
\newcommand{\gep}{$G_{ep}$}
\newcommand{\gmp}{$G_{mp}$}
\newcommand{\rmn}{$r_{mn}$}
\newcommand{\eff}{$\eta(x,y,T_n)$}
\newcommand{\et}{{\em et al.}}

\begin{document}

\begin{frontmatter}


 \title{Precise Neutron Magnetic Form Factors\thanksref{nf}}
 \thanks[nf]{Work supported by the Schweizerische Nationalfonds and Deutsche
Forschungsgemeinschaft, SFB 443}



\author[a]{G.~Kubon}
\author[a]{H.~Anklin}
\author[c]{P.~Bartsch}
\author[c]{D.~Baumann}
\author[c]{W.U.~Boeglin\thanksref{1}}
\thanks[1]{present address: Dept. of Physics, Florida International University, University Park, Miami, FL 33018} 
\author[b]{K.~Bohinc}
\author[c]{R.~B{\"o}hm}
\author[c]{M.O.~Distler}
\author[c]{I.~Ewald}
\author[c]{J.~Friedrich}
\author[c]{J.M~Friedrich}
\author[a]{M.~Hauger}
\author[a]{A.~Honegger} 
\author[c]{P.~Jennewein}
\author[a]{J.~Jourdan\thanksref{5}} 
\thanks[5]{Corresponding author, e-mail: Juerg.Jourdan@unibas.ch}
\author[c]{M.~Kahrau}
\author[c]{K.W.~Krygier} 
\author[c]{A.~Liesenfeld}
\author[c]{H.~Merkel}
\author[c]{U.~M\"uller}
\author[c]{R.~Neuhausen}
\author[a]{Ch.~Normand}
\author[a]{Th.~Petitjean}
\author[c]{Th.~Pospischil}
\author[b]{M.~Potokar}
\author[a]{D.~Rohe}
\author[c]{G.~Rosner\thanksref{2}}
\thanks[2]{present address: Dept. of Physics and Astronomy, University of Glasgow, Glasgow G12 8QQ, Scotland UK} 
\author[c]{H.~Schmieden}
\author[a]{I.~Sick}
\author[b]{S.~\v{S}irca}
\author[a]{Ph.~Trueb}
\author[c]{A.~Wagner}
\author[c]{Th.~Walcher}
\author[a]{G.~Warren}
\author[c]{M.~Weis}
\author[a]{H.~W{\"o}hrle}
\author[a]{M.~Zeier\thanksref{3}}
\thanks[3]{present address: Jefferson Lab., 12000 Jefferson Ave, Newport News, VA 23606} 
\author[a]{J.~Zhao}
\author[a]{B.~Zihlmann\thanksref{4}}
\thanks[4]{present address: NIKHEF, Kruislaan 409, 1098 Amsterdam}

\address[a]{Dept. f\"ur Physik und Astronomie, Universit\"at Basel, Basel, Switzerland}
\address[b]{Institute Jo\v{z}ef Stefan, University of Ljubljana, Ljubljana, Slovenia}
\address[c]{Insitut f\"ur Kernphysik, Johannes Gutenberg--Universit\"at, Mainz, Germany}

\begin{abstract}
Precise data on the neutron magnetic form factor \gmn\ have been 
obtained with measurements of the ratio of cross sections of \deen\
and \deep\ up to momentum transfers of \qsq$= 0.9$~\gev.
Data with typical uncertainties of 1.5\% are presented. These data 
allow for the first time to extract a precise value of the magnetic 
radius of the neutron. 
\end{abstract}


\end{frontmatter}

{\bf Introduction:} \hspace*{0.01cm}
\label{intro}
Detailed information on the inner structure of the nucleon is 
provided by accurate data on the dependence of the 
nucleon form factors on momentum transfer \qsq. 
Such data serve as a sensitive test for models of the nucleon. 
Particularly at low \qsq\ accurate data on the form factors allow for
both a determination of the electromagnetic radii and accurate
calculations of nuclear form factors.

While the proton form factors are known with excellent precision 
over a large range of \qsq,  data for the neutron are of much poorer quality
due to the lack of a free neutron target. This is true for both the
electric form factor, \gen, and to a somewhat lesser extent for the
magnetic one, \gmn. However, with today's high--duty factor, high--current
electron beam facilities and the advances made in 
polarized beam and target technology, significant progress in this area
is being made.

In the past, \gmn\ has been determined mostly from quasi--elastic
$D(e,e')$ cross sections (see references in \cite{Anklin98a}).
The extraction of \gmn\ requires a longitudinal/transverse 
separation and a subtraction of the (dominant) proton magnetic 
contribution. The uncertainties resulting from the deuteron 
wave function, meson exchange currents (MEC), and final state interactions (FSI) 
are greatly amplified by the two subsequent subtractions
and limit the accuracy of \gmn\ from these experiments to $\sim$20\%. 
Due to these limitations alternative techniques have been used in
recent experiments.

One of the techniques determines \gmn\ from the asymmetry
measured in $\vec {^3He}(\vec {e},e')$--scattering \cite{Gao94,Xu00}. 
This challenging technique is presently limited to low \qsq\ where 
today's rigorous non--relativistic 3--body calculations can be applied to
remove the dependence on the nuclear structure, FSI, and MEC \cite{Golak00}.

The neutron magnetic form factor can also be obtained from an exclusive cross section
measurement of $D(e,e'n)$ \cite{Markowitz93}. This technique avoids essentially the 
subtraction of the proton contribution which was responsible for a part
of the large sensitivity to systematic errors in the past. However, the method
still depends on a deuteron model for the extraction of \gmn. 

The best method to minimize the sensitivity to the nuclear structure
is a determination from the ratio 
$R = \sigma (e,e'n) / \sigma (e,e'p) $
on the deuteron in quasi--free kinematics
\cite{Anklin98a,Kubon99,Anklin94,Bruins95}. The ratio is insensitive to 
the deuteron wave function and corrections due to FSI and MEC are 
calculable and small. The price to pay is
the need for a precise measurement of the {\em absolute} efficiency 
$\eta$ of the neutron detector employed.
A measurement of $\eta$ and a detailed study of the detector 
response, however, is possible when using high--intensity neutron beams available 
at the proton--beam facilities \cite{Anklin98a,Anklin94}.

In a pilot experiment it was demonstrated that this method
leads to a determination of \gmn\ with an accuracy of 1.7\% \cite{Anklin94}. 
Similar measurements over an extended \qsq--region are possible with a high--duty
factor electron accelerator like the Mainz Microtron (MAMI) \cite{Herminghaus90}.
Precise measurements of the ratio $R$ were performed in the \qsq--range from 
0.2 to 0.6~\gev\ and values of \gmn\ were extracted with error bars
as low as 1\% \cite{Anklin98a}. In the present work we extend the 
\qsq--range and present data of \gmn\ for \qsq\ of 
0.071, 0.125, 0.359, 0.894 \gev\ in the following labeled as 
kinematics 1 to 4. To check the consistency of such measurements 
the point of the pilot experiment \cite{Anklin94} has been re--measured 
(label 2) using a {\em different} nucleon detector, electron and neutron 
beam facility. 

{\bf Measurements at MAMI:} \hspace*{0.01cm}
At MAMI the yield of \deen\ and \deep\ in quasi--elastic 
kinematics was measured with electron beam energies of 600~MeV and 555~MeV 
for kinematics 1 and 2 and of 855~MeV for kinematics 3 and 4.
A beam current of $0.5~\mu A$ incident on a cylindrical 2~cm thick liquid deuterium 
target cell with 7~$\mu m$ HAVAR windows was employed for kinematics 2 to 4. 
Due to the low proton energy (36~MeV) a target cell 
with a lateral width of only 1~cm liquid deuterium was used for kinematics 1
in order to minimize multiple scattering and energy loss 
effects of the knocked--out protons. Spectrometer A 
\cite{Blomqvist97} with a solid angle of 28~msr detected the 
scattered electron in coincidence with the 
recoiling nucleon. 

The nucleon detector consisted 
of two 10~cm thick plastic converters, $E_f$ and $E_b$, 
preceeded by 3 thin $\Delta E$ counters used
to identify the incident nucleon. The thickness of the $\Delta E$ 
counters was 1.5~mm (5~mm) for kinematics 1 to 2 (3 to 4).
Except for the opening towards the target, the detector was 
shielded with 10~cm thick lead walls. Lead absorbers of 0,1,3, and 20~mm 
thickness were placed at the entrance window of the detector for
kinematics 1 to 4. The absorbers are needed in the \deen\ measurements
in order to absorb low energy photons.
The nucleon detector covered an angular range
of $\pm$78~mr in horizontal as well as vertical direction 
resulting in a solid angle of 24.3~msr. 

The yield ratio has been determined via the {\em simultaneous} measurement
of the \deen\ and the \deep\ reactions which makes the ratio independent  
of the luminosity, dead time effects, and the efficiency of the electron arm. 
Even at the highest \qsq\ a signal--to--noise ratio of 
$\geq 50$ for the \een\ measurement has been achieved.

In the analysis neutrons were defined requiring no hit in at least 2 of the 3
veto counters (${\overline{\Delta E_i}} \cdot \overline{\Delta E_j}$ with $i,j=1,2,3$)
and a hit in $E_f$ in the coincident time window.
The threshold used in the definition of neutrons in $E_f$ was set
at 25~MeV (10~MeV) for kinematics 2 to 4 (1). The number of counted 
neutrons was corrected for the efficiency of the veto condition and 
for misidentified protons due to inefficiencies of the veto counters.
The veto efficiency correction was determined to $<$1\% for 
kinematics 1 and 2, 11.1\% (30.4\%) for kinematics 3 (4). 
The correction for misidentified protons ranged from 0.8\% to 5.2\%. 
The three independent measurements for the number of neutrons, 
determined with the three different veto conditions, 
agreed within~0.16\%(1.5\%) for kinematics 1 to 3 (4). The agreement
of the neutron counts within 1.5\% for kinematics 4 provides confidence on
the validity of the rather large veto correction.

Protons were counted in the TOF spectrum
of $E_f$ with a coincident time signal in at least one of the three
$\Delta E$ counters. The number of protons was corrected
for the inefficiency of the $\Delta E$ counter used.
The final number of protons, obtained using different 
$\Delta E,E_f$ combinations, agreed within 0.11\% for kinematics 1 to 3
and to 0.2\% for kinematics 4. 
The uncertainties in the corrections to the number of protons and
neutrons are small relative to the statistical
error and are included in the error of the yield 
ratios which are summarized in table \ref {rcorr}. 

\noindent {\bf Measurements at PSI:} \hspace*{0.01cm} The 
determination of the neutron detection efficiency was performed
using the high energy neutron beam (100--590~MeV) at the Paul Scherrer 
Institut (PSI) \cite{Daum97} for kinematics 2 to 4 and the monoenergetic 
neutron beam of 68~MeV for kinematics 1 \cite{Henneck87}. 

The high (low) energy neutron beam is produced via the C(p,n) (D(p,n)) 
--reaction with a neutron flux  
of 10$^8$/s (10$^6$/s) for a 5~$\mu A$ proton beam. A 
{\em tagged} high intensity neutron beam was produced 
via the \np\ reaction, scattering the neutrons from a 1~cm thick 
liquid $H_2$--target. The recoil protons were detected with 
$\Delta E_P$ and $E_P$ plastic scintillators recording the amplitude and 
the time--of--flight (TOF). Four multi--wire proportional chambers 
(MWPC) determined the proton trajectory, thereby fixing the 
target coordinates with an accuracy of 
$\pm 1.5$~mm and the recoil angles with an accuracy of $\pm 0.2^o$.

Together with the measured time reference of the radio frequency 
of the cyclotron, the recorded information allowed for a determination 
of the incident neutron energy, the energy of the recoiling proton and its 
recoil angle. This provided a redundant determination 
of the energy and position of the tagged neutron beam which was free 
from contributions of background reactions.

The accuracy of the tagged 
neutron energy from 100 to 490~MeV ranged from 1.9 to 16.6~MeV. 
For 36~MeV tagged neutrons the energy was determined 
to an accuracy of 0.2~MeV. The position on the detector 
surface was determined with an accuracy of $\pm$3.5~mm. 

The nucleon detector was placed in the tagged neutron beam, and its 
absolute efficiency {\em distribution} $\eta(x,y,T_n)$ was measured as a 
function of the point of impact ($x,y$) and the neutron
energy ($T_n$). The knowledge of the distribution $\eta(x,y,T_n)$ 
is necessary because the detector illumination 
is unavoidably different for the $H(n,p)n$ and the \deen\ data taking.
Use of the identical configuration during the ratio measurements 
and the efficiency measurements ensured that absorption 
effects of neutrons were automatically included in the efficiency 
determinations. A more detailed account on the neutron efficiency 
determinations is given in \cite{Fritschi93,Trueb95,Kubon99}.

Considerable care was taken to ensure identical cuts on the energy
deposited in the detectors. $H(e,e'p)$ at MAMI and $H(n,p)$ at PSI  
were used to obtain absolute calibrations of the amplitude spectrum. 
In addition, each converter was monitored for both gain variations 
and baseline shifts 
using two temperature compensated LED's whose light output was 
in turn monitored by very stable PIN-diodes \cite{Anklin94}. 
This information allowed for a determination of the energy scale 
of the converters to an accuracy of 0.8\% during the PSI and the 
MAMI runs. 

Further corrections, relevant in the determination of $R$ were measured
with tagged protons by placing the detector on the recoil arm.
The dominant correction of up to (12.9$\pm$0.7)\% was due to multiple
scattering and energy loss effects leading to proton losses
in the lead absorber. In the same arrangement the variation of 
the light collection efficiency, required to match the
measured \eff\ to the neutron \deen--distribution, was measured. 

Two efficiency measurements bracketed the measurement of the 
yield ratio at MAMI in order to check the reliability of $\eta$.
Consistent results for $\eta$ were found for all kinematic and
absorber conditions.

{\bf Evaluation of R:} \hspace*{0.01cm}
From the measured yield ratios and efficiencies the \deen\ to \deep\ 
cross section ratios $R^{exp}$ were determined with accuracies of 1.7\% to 2.7\%.
The resulting values of $R^{exp}$ are independent of the applied $E_f$ threshold. 
The final results for $R^{exp}$ and the corrections applied in its determination,
all of which are based on measurements, are summarized in table \ref {rcorr}. 

\begin{table}[htb]
\begin{center}
\begin{tabular}{ccccrrrr}
\hline
 & & & & & & & \\ [-4mm]
\multicolumn{4}{l}{\qsq\ \gev} & 0.071 & 0.125 
& 0.359 & 0.894 \\ [2mm]
\hline
 & & & & & & & \\ [-4mm]
\multicolumn{4}{l}{Yield ratio} & $\pm$0.5 & 
$\pm$1.0 & $\pm$0.9 & $\pm$2.1 \\ 
\multicolumn{4}{l}{Illumin. matching} & $\pm$1.0 &
$\pm$0.5 & $\pm$0.7  & $\pm$0.6 \\
\multicolumn{4}{l}{Thresh. calibration} & $\pm$0.8 &
$\pm$0.8 & $\pm$0.8  & $\pm$0.8 \\
\multicolumn{4}{l}{Proton losses} & $-7.6\pm$0.9 &
$-11.7\pm$1.0 & $-8.7\pm$0.7  & $-12.1\pm$0.7 \\
\multicolumn{4}{l}{(p,n)--correction} & $-0.00\pm$0.02 &
$-0.07\pm$0.02 & $-0.3\pm$0.1  & $-1.9\pm$0.6 \\
\multicolumn{4}{l}{Error of $\eta$} & $\pm$0.6 &
$\pm$1.0 & $\pm$0.5 & $\pm$0.8 \\
\multicolumn{4}{l}{$T_n$ and $x,y$} & $\pm$0.4 &
$\pm$1.1 & $\pm$0.6 & $\pm$0.8 \\ [2mm]
\hline
 & & & & & & & \\ [-4mm]
\multicolumn{4}{l}{$R$} & 0.0973 & 0.139 & 0.279 & 0.311 \\
\multicolumn{4}{l}{Rel. error of $R$ in \%} & $\pm$1.7  & 
$\pm$2.0 & $\pm$1.7 & $\pm$2.7 \\ [2mm]
\hline \\ [-4mm]
\end{tabular}
\parbox{12cm}{\caption{\label{rcorr} Results for $R$ and the experimentally determined 
errors and corrections in \% of $R$. Except for the yield ratio and $\eta$
the errors are mainly systematic in nature.}}
\end{center}
\end{table}

To extract \gmn\ from the measured $R^{exp}$ values we have to 
take into account the effects of FSI, MEC, and isobar currents (IC)
beyond the plane wave impulse approximation (PWIA).
To this end, we used theoretical results of $R^{theo}$
calculated with the Paris potential by Arenh\"ovel \cite{Arenhoevel79}.
The total corrections are listed in table \ref{results} as a 
deviation $D = (R_{PWIA}/R^{theo}) - 1$ from the PWIA--value.
The dominant contribution to $D$ (up to 99\%) 
is due to FSI, mostly 
charge exchange scattering. The contributions from MEC 
and IC are of order 0.5\%, and relativistic effects are negligible\cite{Arenhoevel97}.

At \qsq~=~0.071 \gev\ where the correction is largest
the dependence of $D$ on the nucleon--nucleon potential was studied.
A variation of typically 0.9\% in $D$ was found when using the Bonn--R--space, 
the Argonne~V14, and the Paris potential. This is well within the relative systematic 
uncertainties (FSI: $\pm$8\%, MEC: $\pm$40\%, IC: $\pm$60\%) used to
calculate the errors of table \ref {results}.

The resulting value $R^{exp}_{PWIA} = (\sigma_{e-n}/\sigma_{e-p})_{PWIA} =
R^{exp} \cdot (1 + D)$ is the experimental ratio of the $e-n$ cross section
(which is essentially given by \gmn) to the $e-p$ cross 
section corrected for non--PWIA contributions. The contribution of 
the neutron electric form factor to $R^{exp}_{PWIA}$ is small; it introduces only a 
small additional uncertainty despite the 
poor knowledge of \gen\ (see table \ref{results}). 

\begin{table}[hbt]
\begin{center}
\begin{tabular}{cccccrrrr}
\hline
 & & & & & & & & \\ [-4mm]
\multicolumn{5}{l}{\qsq\ \gev} & 0.071 & 0.125 & 0.359 & 0.894 \\ [2mm]
\hline
 & & & & & & & & \\ [-4mm]
\multicolumn{5}{l}{$D$ (\%)}& 
$-24.1\pm2.0$ & $-9.9\pm0.9$ & $-3.7\pm0.4$ & $-1.2\pm0.2$\\ 
\multicolumn{5}{l}{Contr. of \gen$~$(\%)}&
$-1.4\pm0.3$ & $-1.9\pm0.4$ & $-2.8\pm0.6$ & $-0.6\pm0.1$\\ 
\multicolumn{5}{l}{$\sigma_{e-p}/\sigma_D$} & 
$0.948\pm0.009$ & $0.931\pm0.014$ & $0.933\pm0.019$ & $1.089\pm0.023$ \\ 
\multicolumn{5}{l}{$G_{mn}/(\mu_nG_D)$}& 
$0.990\pm0.013$ & $0.967\pm0.013$ & $0.989\pm0.014$ & $1.062\pm0.017$ \\ [2mm]
\hline \\ [-4mm]
\end{tabular}
\parbox{12.5cm}{\caption{\label{results} Results for \gmn\ relative to the 
dipole form factor $G_D=(1+Q^2/0.710)^{-2}$. The error on \gmn\ includes
both the experimental contribution (table \protect{\ref {rcorr}}, 
$\sigma_{e-p}$, \gen) and the ones due to theory.}}
\end{center}
\end{table}

To evaluate the $e-p$ cross section, $\sigma_{e-p}$, 
we used the world's supply of {$\sigma_{e-p}$ data}
in a range of 0.5$~fm^{-1}$ around the desired $Q$. 
In this range of $Q$, we used a parameterization with a relative 
$Q$--dependence as given by the Mergell \et\ fit \cite{Mergell96}
to \gep\ and \gmp,
with the overall normalizations fitted to the world data. 
The statistical errors of the data have been treated in the 
standard way, while the systematic errors have been accounted for by 
changing the data of each set by its error, refitting, and adding the 
changes due to systematic errors in quadrature. The resulting $e-p$ 
cross sections relative to the ones obtained with dipole form factors
are listed in table \ref{results} together with the final results for \gmn. 

\begin{figure}[htb]
\begin{center}
\includegraphics[scale=0.6]{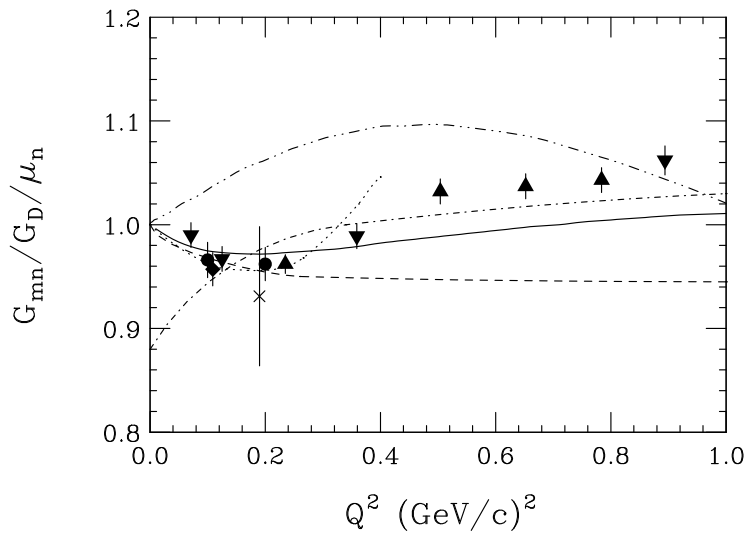}
\parbox{13cm}{\caption[]{
The figure shows the present results ($\blacktriangledown$),
together with results of the previous MAMI/PSI \protect{\cite{Anklin98a} ($\blacktriangle$)}, 
the NIKHEF/PSI \protect{\cite{Anklin94} ($\blacklozenge$)}, 
the Bates $^3\vec{\mathrm{He}}$ \protect{\cite{Gao94} ($\times$)}, 
and the JLab $^3\vec{\mathrm{He}}$ experiments 
\protect{\cite{Xu00}} ($\bullet$), 
in comparison to various model calculations. 
Solid: Mergell \et\ {\protect\cite{Mergell96}},
dot: Kubis {\protect\cite{Kubis00}},
dash--dot: Eich {\protect\cite{Eich88}},
dash: Schlumpf {\protect\cite{Schlumpf94}},
dash--dot--dot: Lu \et\ {\protect\cite{Lu97}}.
}\label{gmn_letter}} 
\end{center} 
\end{figure}

The present data on \gmn\ are shown in figure \ref{gmn_letter} together 
with the recently determined data sets ($>1990$) of refs 
\cite{Anklin98a,Gao94,Xu00,Anklin94}
and some recent calculations. The present data extend the previously
investigated \qsq--region in 
both directions and allow for a direct comparison of the data measured 
at NIKHEF/PSI \cite{Anklin94} and JLab \cite{Xu00}. The agreement
of these data measured with different techniques at different
facilities at an unprecedented level of precision is very satisfactory. 

On the other hand, these combined data do not agree with the
measurements by \cite{Markowitz93,Bruins95} (not shown in figure
\ref{gmn_letter}). This is due to the fact that in these experiments 
the three--body reactions $D(e,p)ne'$ and $H(e,\pi)ne'$ were used to 
tag the recoiling neutron in the $\eta$--determination. This, however,
would have required significant corrections for neutrons that miss
the nucleon detector; these corrections were not applied \cite{Jourdan97}.

The data shown in figure \ref{gmn_letter} clearly
differ from the crude empirical expression $G_D=(1+Q^2/m^2)^{-2}$ with
$m^2 = 0.71$ \gev\
used to remove the dominant \qsq--dependence in table \protect{\ref {results}}
and figure \protect{\ref {gmn_letter}}. 
In addition, the data show significant differences to both the 
non--relativistic constituent
quark model calculation by Eich \cite{Eich88}, and the relativistic 
version by Schlumpf \cite{Schlumpf94}. Similar differences are observed
when comparing the data to a recent cloudy bag model calculation by Lu 
\et\ \cite{Lu97}. The data are also compared to results of the relativistic 
chiral perturbation theory by Kubis and Meissner \cite{Kubis00} 
and the calculation by Mergell \et\ \cite{Mergell96} based on a fit of the 
proton data using dispersion theoretical arguments. While none of these 
calculations describe the data satisfactorily the tendency is given by 
the calculation of Ref. \cite{Mergell96}.

{\bf Parameterization of \gmn:} \hspace*{0.01cm}
The present data allow for the first time a purely experimental extraction
of the root--mean--square (rms) magnetic radius of the neutron defined as:
\begin{equation}
<r_{mn}^2> = -6\left(\frac{1}{\mu_n}\frac{dG_{mn}(Q^2)}{dQ^2}\right)_{Q^2=0}
\end{equation}
In the past, the experimental information on \rmn\ was based on a
dispersion--theoretical analysis of the combined set of electromagnetic 
form factors of both neutron and proton. In this 
framework \rmn\ is determined mostly from constraints other than the 
experimental \gmn\ data which up to now were very limited in accuracy.
  
The present determination of \rmn\ uses {\em only} the experimental data 
on \gmn. The data of figure \ref{gmn_letter} together with the higher 
\qsq\ data from \cite{Lung93} are taken into account. 
A continued fraction parameterization given by:
\begin{equation}
 G_{mn}(Q^2) = \frac{\mu_n}{\displaystyle 1
                    + \frac{Q^2b_1}{\displaystyle 1
		       + \frac{Q^2b_2}{\displaystyle 1
		          + \cdots}}}
\end{equation}			  
is fitted to the data. The magnetic radius of the neutron
is related to the parameter $b_1$ via $<r_{mn}^2> = 6\cdot b_1$.  

The continued fraction representation \cite{Klarsfeld86} has the advantage 
to converge in a wider domain than the usual power series 
expansion. One finds, for example, that five terms are sufficient to 
reproduce \gmn\ of the dispersion analysis by Hoehler \cite{Hoehler76} or
Mergell \cite{Mergell96} for \qsq\ up to 4~\gev.

\begin{figure}[hbt]
\begin{center}
\includegraphics[scale=0.6]{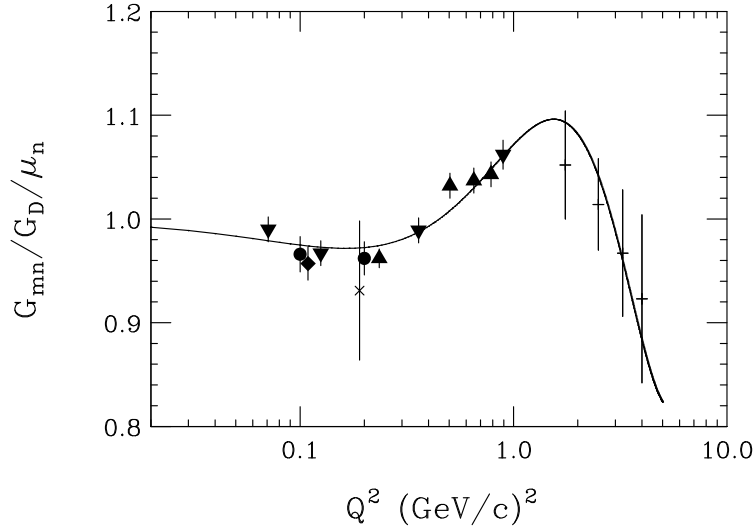}
\parbox{13cm}{\caption[]{
The figure shows the continued fraction fit to the data. Symbols 
for the data as in figure \ref{gmn_letter}) 
plus the data by Lung \et\ ($+$)
\protect{\cite{Lung93}}.
}\label{gmn_r}} 
\end{center} 
\end{figure}
Figure \ref{gmn_r} shows the result of the fit using the five terms
$b_1,..b_5$ = 3.26, $-0.272$, $0.0123$, $-2.52$, $2.55$ $\mathrm{(GeV/c)}^{-2}$ 
with a $\chi_{red} = 0.91$.
For the magnetic radius we find $r_{mn} = 0.873\pm 0.011$~\fm. 
The uncertainty of 1.1\%, an improvement of a factor of 10
compared to previous determinations, covers 
the error of the fit (0.6\%) and a systematic uncertainty due to the
choice of the fit function. The latter was determined by
fitting with different parameterizations pseudo data from a dispersion analysis placed at the
\qsq--values and with the errors of the real data. 


The present result provides an additional observable to test
theoretical calculations of the nucleon.
The calculation by Buchmann \et\ \cite{Buchmann94} has been performed 
in the framework of a constituent quark model including
gluon, meson, and confinement exchange currents. The calculation
predicts a radius of 0.891~\fm\ in marginal agreement with the present
experimental result. The agreement is worse when comparing
to the very recent result of 0.84~\fm\ by Kubis and Meissner 
\cite{Kubis00} obtained in fourth order relativistic baryon chiral 
perturbation theory. 

{\bf Summary:} \hspace*{0.01cm} The present data combined with the other data
of figure \ref{gmn_letter} have improved our knowledge 
of the neutron magnetic form factor in the \qsq--range from 0.07 to
0.89 \gev\ by a factor of 10 compared to determinations based on
quasi--elastic $(e,e')$--data performed in  the past \cite{Anklin98a}.
The improvement is mostly due to the fact that 
the exploitation of the ratio $R = \sigma (e,e'n) / \sigma (e,e'p)$
depends least on the input of theory; the measurement of $R$ becomes possible by
performing the needed calibrations using a high--intensity tagged neutron 
beam, and by using an electron beam with 100\% duty factor. 
The increase of accuracy allows for a detailed comparison to 
theoretical model calculations.
The agreement of three high--precision experiments gives us confidence 
in the reliability of the gained knowledge on the magnetic properties of the 
neutron. 






\begin{thebibliography}{10}

\bibitem{Anklin98a}
H.~Anklin {\em et al.}
\newblock {\em Phys. Lett. B}, 428:248, 1998.

\bibitem{Gao94}
H.~Gao {\em et al.}
\newblock {\em Phys. Rev. C}, 50:R546, 1994.

\bibitem{Xu00}
W.~Xu {\em et al.}
\newblock {\em Phys. Rev. Lett.}, 85:2900, 2000.

\bibitem{Golak00}
J.~Golak, G.~Ziemer, H.~Kamada, H.~Witala, and W.~Gl{\"o}ckle.
\newblock {\em Phys. Rev. C}, 63:034006, 2001.

\bibitem{Markowitz93}
P.~Markowitz {\em et al.}
\newblock {\em Phys. Rev. C}, 48:R5, 1993.

\bibitem{Kubon99}
G.P. Kubon.
\newblock PhD thesis, University of Basel, 1999.

\bibitem{Anklin94}
H.~Anklin {\em et al.}
\newblock {\em Phys. Lett. B}, 336:313, 1994.

\bibitem{Bruins95}
E.E.W. Bruins {\em et al.}
\newblock {\em Phys. Rev. Lett.}, 75:21, 1995.

\bibitem{Herminghaus90}
H.~Herminghaus, H.~Euteneuer, and K.H. Kaiser.
\newblock Proc. LINAC'90, Albuquerque, New Mexico, USA, p362, 1990.

\bibitem{Blomqvist97}
K.I.~Blomqvist {\em et al.}
\newblock {\em Nucl. Instr. and Methods A}, 403:263, 1998.

\bibitem{Daum97}
J.~Arnold {\em et al.}
\newblock {\em Nucl. Instr. and Meth. A}, 386:211, 1997.

\bibitem{Henneck87}
R.~Henneck {\em et al.}
\newblock {\em Nucl. Instr. Meth. A}, 259:329, 1987.

\bibitem{Fritschi93}
D.~Fritschi.
\newblock {\em PhD thesis}, Basel University, 1993.

\bibitem{Trueb95}
P.~Trueb.
\newblock {\em PhD thesis}, Basel University, 1995.

\bibitem{Arenhoevel79}
W.~Fabian and H.~Arenh\"ovel.
\newblock {\em Nucl. Phys. A}, 314:253, 1979.

\bibitem{Arenhoevel97}
F.~Ritz, H.~G\"oller, T.~Wilbois, and H.~Arenh\"ovel.
\newblock {\em Phys. Rev. C}, 55:2214, 1997.
G.~Beck, T.~Wilbois, and H.~Arenh\"ovel.
\newblock {\em Few--Body Syst.}, 15:39, 1993.

\bibitem{Mergell96}
P.~Mergell, U.-G. Meissner, and D.~Drechsel.
\newblock {\em Nucl. Phys. A}, 596:367, 1996.

\bibitem{Kubis00}
B.~Kubis and U.-G. Meissner.
\newblock {\em Nucl. Phys. A}, 679:3, 2001.

\bibitem{Eich88}
E.~Eich.
\newblock {\em Z. Phys. C}, 45:627, 1988.

\bibitem{Schlumpf94}
F.~Schlumpf.
\newblock {\em J. Phys. G}, 20:237, 1994.

\bibitem{Lu97}
D.H. Lu, A.W. Thomas, and A.G. Williams.
\newblock {\em Phys. Rev. C}, 57:2628, 1997.

\bibitem{Jourdan97}
J.~Jourdan, I.~Sick, and J.~Zhao.
\newblock {\em Phys. Rev. Lett.}, 79:5186, 1997.

\bibitem{Lung93}
A.~Lung {\em et al.}
\newblock {\em Phys. Rev. Lett.}, 70:718, 1993.

\bibitem{Klarsfeld86}
S.~Klarsfeld, J.~Martorell, J.A. Oteo, M.~Nishimura, and D.W.L. Sprung.
\newblock {\em Nucl. Phys. A}, 456:373, 1986.

\bibitem{Hoehler76}
G.~Hoehler, E.~Pietarinen, I.~Sabba-Stefanescu, F.~Borkowski, G.G. Simon, V.H.
  Walther, and R.D. Wendling.
\newblock {\em Nucl. Phys. B}, 114:505, 1976.

\bibitem{Buchmann94}
A.~Buchmann, E.~Hernandez, and K.~Yazaki.
\newblock {\em Nucl. Phys. A}, 569:661, 1994.

\end{thebibliography}

\end{document}